\PassOptionsToPackage{dvipsnames}{xcolor}
\documentclass[12pt,a4paper]{article}
\usepackage{amsmath,amssymb,url}
\usepackage{orcidlink}
\usepackage{graphicx,tabularx}
\usepackage{array,dsfont}
\usepackage{thophys}
\usepackage[english]{babel}
\usepackage[T1]{fontenc}
\hypersetup{colorlinks=true,linkcolor=ForestGreen,citecolor=ForestGreen,filecolor=ForestGreen,urlcolor=ForestGreen}

\makeatletter
\newif\if@preliminary
\@preliminaryfalse
\def\preliminary{\@preliminarytrue}
\def\preprintno#1{\def\@preprintno{#1}}
\def\address#1{\def\@address{#1}}
\def\email#1#2{\thanks{\tt #1@{}#2}}
\def\abstract#1{\def\@abstract{#1}}
\renewcommand\abstractname{ABSTRACT}
\newlength\preprintnoskip
\newlength\abstractwidth
\@titlepagetrue
\renewcommand\maketitle{\begin{titlepage}%
  \let\footnotesize\small
  \hfill\parbox{\preprintnoskip}{%
  \begin{flushright}\@preprintno\end{flushright}}\hspace*{.5cm}
  \vskip 60\p@
  \begin{center}%
    {\Large\bf\boldmath \@title \par}\vskip 1cm%
    {\sc\@author \par}\vskip 3mm%
    {\@address \par}%
    \if@preliminary
      \vskip 2cm {\large\sf PRELIMINARY DRAFT \par \@date}%
    \fi
  \end{center}\par
  \@thanks
  \vfill
  \begin{center}%
    \parbox{\abstractwidth}{\centerline{\abstractname}%
    \vskip 3mm%
    \@abstract}
  \end{center}
  \end{titlepage}%
  \setcounter{footnote}{0}%
  \let\thanks\relax\let\maketitle\relax
  \gdef\@thanks{}\gdef\@author{}\gdef\@address{}%
  \gdef\@title{}\gdef\@abstract{}\gdef\@preprintno{}
}%
\def\@citex[#1]#2{\if@filesw\immediate\write\@auxout{\string\citation{#2}}\fi
  \def\@citea{}\@cite{\@for\@citeb:=#2\do
    {\@citea\def\@citea{,\penalty\@m}\@ifundefined
       {b@\@citeb}{{\bf ?}\@warning
       {Citation `\@citeb' on page \thepage \space undefined}}%
\hbox{\csname b@\@citeb\endcsname}}}{#1}}
\def\citerange{\@ifnextchar [{\@tempswatrue\@citexr}{\@tempswafalse\@citexr[]}}
\def\@citexr[#1]#2{\if@filesw\immediate\write\@auxout{\string\citation{#2}}\fi
  \def\@citea{}\@cite{\@for\@citeb:=#2\do
    {\@citea\def\@citea{--\penalty\@m}\@ifundefined
       {b@\@citeb}{{\bf ?}\@warning
       {Citation `\@citeb' on page \thepage \space undefined}}%
\hbox{\csname b@\@citeb\endcsname}}}{#1}}
\long\def\@makecaption#1#2{%
  \sbox\@tempboxa{#1: \emph{#2}}%
  \ifdim \wd\@tempboxa >\hsize
    #1: \emph{#2}\par
  \else
    \hbox to\hsize{\hfil\box\@tempboxa\hfil}%
  \fi
  \vskip\belowcaptionskip}
\def\fmslash{\@ifnextchar[{\fmsl@sh}{\fmsl@sh[0mu]}}
\def\fmsl@sh[#1]#2{%
  \mathchoice
    {\@fmsl@sh\displaystyle{#1}{#2}}%
    {\@fmsl@sh\textstyle{#1}{#2}}%
    {\@fmsl@sh\scriptstyle{#1}{#2}}%
    {\@fmsl@sh\scriptscriptstyle{#1}{#2}}}
\def\@fmsl@sh#1#2#3{\m@th\ooalign{$\hfil#1\mkern#2/\hfil$\crcr$#1#3$}}
\makeatother

\newcommand\ltap{\
  \raise.3ex\hbox{$<$\kern-.75em\lower1ex\hbox{$\sim$}}\ }
\newcommand\gtap{\
  \raise.3ex\hbox{$>$\kern-.75em\lower1ex\hbox{$\sim$}}\ }

\newcommand\simge{\mathrel{%
   \rlap{\raise 0.511ex \hbox{$>$}}{\lower 0.511ex \hbox{$\sim$}}}}
\newcommand\simle{\mathrel{
   \rlap{\raise 0.511ex \hbox{$<$}}{\lower 0.511ex \hbox{$\sim$}}}}

\newcommand\be{\begin{equation}}
\newcommand\ee{\end{equation}}
\newcommand\bea{\begin{eqnarray}}
\newcommand\eea{\end{eqnarray}}
\newcommand\ba{\begin{array}}
\newcommand\ea{\end{array}}

\def\bq{\begin{equation}}
\def\eq{\end{equation}}
\def\ba{\begin{eqnarray}}
\def\ea{\end{eqnarray}}

\begin{document}

\date{\today}

\preprintno{DESY-23-108, P3H-23-049, SI-HEP-2023-17}

\title{New developments on the WHIZARD event generator}

\author{\underline{J\"urgen
  Reuter}\email{juergen.reuter}{desy.de}$^{\ast a\,\orcidlink{0000-0003-1866-0157}}$,  
  Pia Bredt\email{pia.bredt}{uni-siegen.de}$^{b\,\orcidlink{0000-0003-4579-0387}}$,
  Wolfgang
  Kilian\email{kilian}{physik.uni-siegen.de}$^{b\,\orcidlink{0000-0001-5521-5277}}$,
  Maximilian L\"oschner\email{maximilian.loeschner}{desy.de}$^{a\,\orcidlink{0000-0002-0850-257X}}$, 
  Krzysztof
  M\k{e}ka{\l}a\email{krzysztof.mekala}{desy.de}$^{a,c\,\orcidlink{0000-0003-4268-508X}}$,
  Thorsten
  Ohl\email{ohl}{physik.uni-wuerzburg.de}$^{d\,\orcidlink{0000-0002-7526-2975}}$, 
  Tobias Striegl\email{striegl}{physik.uni-siegen.de}$^b$,
  Aleksander
  Filip \.Zarnecki\email{zarnecki}{fuw.edu.pl}$^{c\,\orcidlink{0000-0001-8975-9483}}$  
}

\address{\it%
$^a$Deutsches Elektronen-Synchrotron DESY, 
  Notkestr. 85, 22607 Hamburg, Germany
\\[.5\baselineskip]
$^b$
Department of Physics, University of Siegen, Walter-Flex-Stra{\ss}e 3, 57068 Siegen, Germany
\\[.5\baselineskip]
$^c$
Faculty of Physics, University of Warsaw,
Pasteura 5, Warszawa, 02-093, Poland
\\[.5\baselineskip]
$^d$
University of W\"urzburg, Institut f\"ur Theoretische Physik und
Astrophysik, Emil-Hilb-Weg 22, 97074 W\"urzburg, Germany
\\[.8\baselineskip]
\mbox{}  $^\ast$ {\em Speaker, Talk presented at the International Workshop on Future Linear Colliders (LCWS 2023), 15-19 May 2023. C23-05-15.3.}
}

\abstract{
  We give a status report on new developments in the WHIZARD event
  generator, including NLO electroweak automation for $e^+e^-$ colliders,
  loop-induced processes, POWHEG matching, new features in the UFO
  interface and the current development for matching between exclusive
  photon radiation and fixed-order LO/NLO electroweak (EW)
  corrections. We report on several bug fixes relevant for certain
  aspects of the ILC250 Monte Carlo (MC) mass production, especially
  on the normalization of matching EPA samples 
  with full-matrix element samples. Finally, we mention some ongoing
  work on efficiency improvements regarding parallelization of matrix
  elements and phase space sampling, as well as plans to revive the
  top threshold simulation.
}

\maketitle


\section{Introduction}

This is an update summarizing the status report on the {\sc Whizard}
event generator given at the LCWS 2023 conference at SLAC in May this
year. The update comprises current development on the automation of
higher-order corrections with a focus on linear electron-positron, and
in general, lepton colliders, within the last few years since the last
pre-Corona meeting of the linear collider community in Sendai 2019.

{\sc Whizard}~\cite{Kilian:2007gr} is a multi-purpose Monte Carlo
event generator framework that apart from hadronization and a few
minor topics covers all aspects of collider simulations, at hadron,
electron-positron, lepton-hadron, photon and, more recently, also muon
colliders. It is based upon its tree-level matrix-element generator
{\sc O'Mega}~\cite{Moretti:2001zz,Reuter:2002gn}, using functional
methods to recursively construct amplitudes~\cite{Ohl:2023bvv}. For
QCD, the color-flow formalism is used~\cite{Kilian:2012pz}. The {\sc
Whizard} framework includes tools for linear lepton and photon
collider beam simulation ({\sc Circe}
package~\cite{Ohl:1996fi}). Phase space integration and MC simulation
is done with an adaptive multi-channel approach~\cite{Ohl:1998jn},
that is also available in the form of an MPI-based
parallelization~\cite{Brass:2018xbv}.

The most substantial progress, underlined in the release of versions
v3.0 in April 2021 and v3.1 in December 2022 was the complete
automation of next-to-leading order (NLO) QCD and electroweak (EW)
corrections in the SM for both lepton and hadron colliders, the
support for loop-induced processes and the quasi-automatic NLO
matching to parton showers via the POWHEG algorithm. This is
documented
in~\cite{Weiss:2017qbj,ChokoufeNejad:2017rag,Rothe:2021sml,Stienemeier:2022wmy,Bredt:2022nkq}, 
and will be detailed in Sec.~\ref{sec:prec}. In Sec.~\ref{sec:bsm} we
will discuss the second milestone, the completion of the UFO interface
for BSM models, together with technical details that were important for
the ILC MC mass production for the 250 GeV energy stage. Finally, in
Sec.~\ref{sec:outlook} we give an outlook on open projects and plans
for the future. 


\section{NLO automation, NLO matching and precision simulations}
\label{sec:prec}

{\sc Whizard} provides a complete automation of NLO SM (QCD and EW)
corrections for lepton and hadron colliders. These build upon earlier
work for NLO QCD~\cite{Binoth:2009rv,Greiner:2011mp} and NLO
EW~\cite{Kilian:2006cj,Robens:2008sa}, that were using
process-specific Catani-Seymour subtraction~\cite{Catani:1996vz} or
phase-space slicing, respectively. The modern completely general,
process-independent implementation of NLO corrections is based on the
FKS subtraction formalism~\cite{Frixione:1995ms,Frixione:1997np}. This
implementation has been first applied to off-shell top-pair and
$t\bar{t}$-associated Higgs production at lepton
colliders~\cite{ChokoufeNejad:2016qux}, also matched to NRQCD
threshold resummation at NLL level~\cite{Bach:2017ggt}. This
implementation released for v2.4.1 in 2017 has been recently
revalidated in v3.0, and there was an improvement in the treatment of
the p-wave NLL formfactor in the matched calculation. We have started
to interface the matched NLO event generation to a fast detector simulation,
however, there are still a few technical issues to be solved: the full
setup for beamstrahlung, polarization, QED ISR and top threshold
matching does not yet work, and the default setup (plain beams or with
just QED ISR) suffers from a very high number of negative weights. On
the other hand, the POWHEG matching (cf. below) works technically with
the top threshold setup.

\begin{figure}
\includegraphics[width=.49\textwidth]{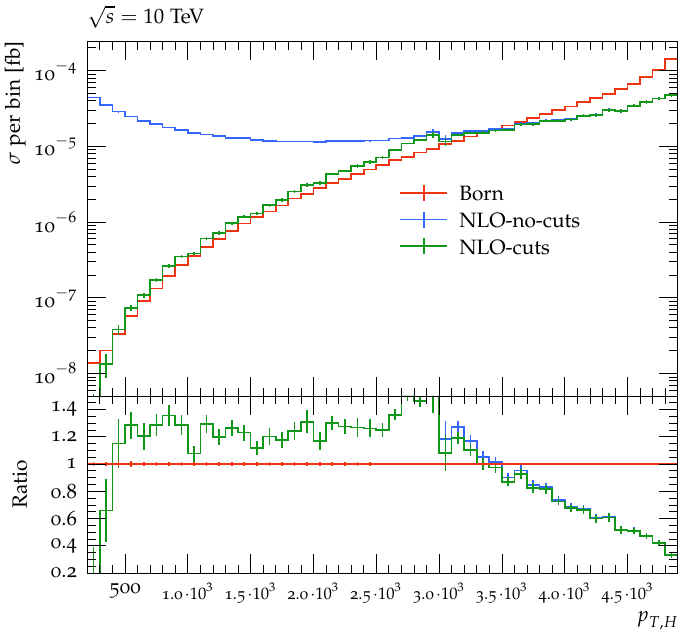}
\includegraphics[width=.49\textwidth]{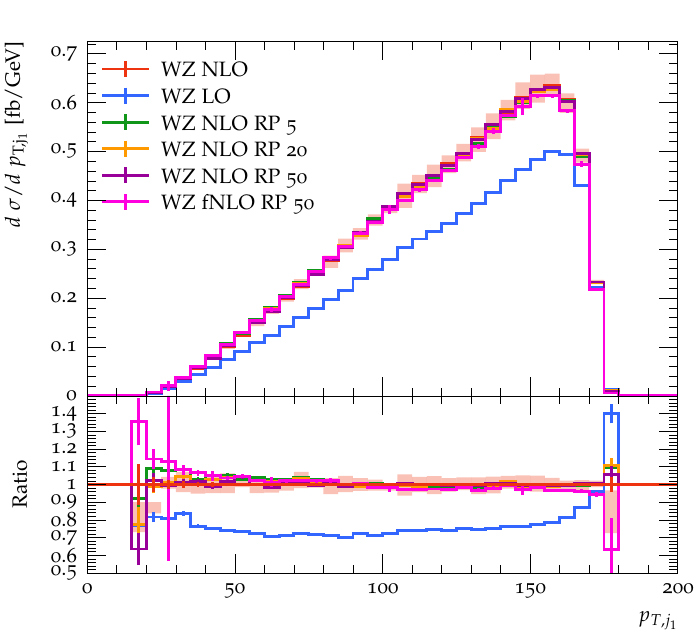}
\caption{Examples for NLO functionality in {\sc Whizard}: differential
$p_T$ distribution for the process $\mu^+\mu^- \to ZH$ at NLO EW at a
10 TeV muon collider (left), POWHEG matching for the process
$e^+e^- \to t\bar{t}j$ at a 1 TeV ILC with the $p_T$ distribution of
the hardest jet.}
\label{fig:nlo_plots}
\end{figure}

Fixed-order NLO QCD processes have been
validated both for LHC, $e^+e^-$ colliders and muon colliders and have
been presented
e.g. in~\cite{Ballestrero:2018anz,Stienemeier:2021cse,Campbell:2022qmc}. 
For the NLO processes, {\sc Whizard} uses the BLHA
interface~\cite{Binoth:2010xt,Alioli:2013nda}  and
supports as one-loop providers (OLPs) {\sc
GoSam}~\cite{GoSam:2014iqq}, {\sc
OpenLoops}~\cite{Cascioli:2011va,Buccioni:2019sur} and  
{\sc Recola}~\cite{Actis:2016mpe}. Independent variations of
factorization and renormalization scale are possible. With v3.1 of
{\sc Whizard}, also loop-induced processes are available. For
fixed-order NLO EW corrections and mixed QCD-EW corrections, extensive
validations have been done for many classes of LHC processes (EW
bosons, top pairs, VBF Higgs, single top etc., which necessitates both
jet clustering and photon
recombination)~\cite{Bredt:2022nkq}. Recently, also NLO EW 
corrections for lepton collisions with massive initial states have
been automated~\cite{Bredt:2022dmm}. The most crucial ingredient here 
are numerically stable eikonal terms for soft subtraction. An example
differential distribution, the Higgs $p_T$ is shown for
$\mu^+\mu^- \to ZH$ at a 10 TeV muon collider at NLO EW,
Fig.~\ref{fig:nlo_plots} (left).  For massless leptons in the initial
state, collinear factorization can be used to define lepton PDFs,
either at leading-logarithmic (LL)
accuracy~\cite{Skrzypek:1990qs,Cacciari:1992pz} or at NLL 
accuracy~\cite{Frixione:2019lga,Bertone:2019hks}. The LL lepton PDFs
have been implemented in {\sc Whizard} from the very beginning of
v1.x (together with the photon content in the form of
equivalent-photon approximation), while NLL QED PDFs have been
implemented and validated for v3.0.3. At the moment, they are being
combined with the initial-state FKS subtraction automation. In the
past 2-3 years there has been also improvement on the proper matching
both for photons from ISR (with a semi-inclusive description of a
single photon per beam with a QED-inspired transverse momentum
generation) with matrix-element photons~\cite{Kalinowski:2020lhp} as
well as between EPA $e^+e^- \to \gamma\gamma \to X$ processes and full
processes $e^+e^- \to e^+e^- X$.

In~\cite{Stienemeier:2022wmy}, a completely automated NLO matching
using the POWHEG method~\cite{Frixione:2007vw} has been presented
which was released with v3.1 of {\sc Whizard}. This can be applied for
NLO QCD corrections at hadron and lepton colliders, both for massless
and massive emitters. Fig.~\ref{fig:nlo_plots} shows on the right as
an example the $p_T$ distribution of the hardest jet for the process
$e^+e^- \to t\bar{t}j+X$ at a 1 TeV ILC with the LO distribution in
blue, the NLO distribution in red and the other colors show
POWHEG-matched curves for different parameters of the real
partitioning. A generalization of POWHEG matching to NLO QED/EW is
conceptually straightforward, and will be one of the next steps.

For beam simulations, the spectrum generator {\sc Circe2} now allows
also for muons as beam particles. However, for the muon collider, the
Cool Copper Collider (CCC) as well as for the photon collider based on
the copper technology, XCC, are still pending, while the new spectra
for FCC-ee (with four interaction points) will soon be available.


\section{BSM models, UFO interface and technical details}
\label{sec:bsm}

The release of v3.0 of {\sc Whizard} consituted also the completion of
the general UFO~\cite{Degrande:2011ua} interface, which allows to
import arbitrary BSM models via the UFO interface into {\sc
Whizard}. Before that, BSM models were for a very long time
hard-coded, e.g. supersymmetric models like the (N)MSSM, Little Higgs
models, or extra-dimensional models, or to be ingested via dedicated
interfaces~\cite{Christensen:2010wz}. {\sc Whizard/O'Mega} comes with
a dedicated parser that allows to transfer the Python-based UFO model
definition into models that can be understood by {\sc Whizard} and can
then be used in the same way as hard-coded models. They do support
particles of spin 0, 1/2, 1, 3/2 and 2, the basic color
representations (the implementation of exotic color representationsand
totally antisymmetric color couplings will be released soon, after the
required external interface definitions have been finalized), both
fermion-number preserving and violating models (for a first application,
cf.~\cite{Mekala:2022cmm,Mekala:2023diu}), higher-multiplicity
vertices (cf.~\cite{Han:2021lnp}), and they support 
non-standard propagators as 
well. In fact, several of the extensions that have been summarized
recently in~\cite{Darme:2023jdn} are supported in {\sc Whizard} v3.1.
{\sc Whizard} also allows to read in a (SUSY) Les Houches-type input
file that could have been generated from the spectrum generator,
e.g. for parameter scans in BSM models. 
Ongoing work on the interface will soon allow for specification of
exclusive coupling orders, both for SM (QCD vs. QED/EW) and for BSM,
e.g. in operator insertions. In a next step, this will then facilitate
the bookkeeping for the definition of several components
for NLO calculations, e.g. color- and spin-correlated matrix elements
used in the subtraction.

On the technical side, besides the very powerful parallelized adaptive
MC integration based on the Message-Processing Interface (MPI, where
{\sc Whizard} now comes with two different implementations, a
"classic" one and one with a load balancer for non-blocking
communication), there is also an implementation where {\sc O'Mega}
matrix elements are generated as CUDA libraries and are then evaluated
on the GPU. The speed-ups seen there are expected from a careful
profiling of the parallel and serial components of the program. As
next steps, we will test this for NLO matrix elements where the
evaluation per phase-space point is even more costly, and we also plan
to construct the phase space itself on the GPU. 

The release of v3.x was also used to provide some technical features
and to fix a lot of technical bugs. {\sc Whizard} v3.x now comes with
an API that allows to call the program as a library, e.g. for
single-event generation from a detector-level simulation
tool. Using this API, examples are provided for calling {\sc Whizard}
via C, C++, Fortran and Python. The new feature also comprise support
for the new Darwin OS and the new Apple M1/M2 processors. We are
grateful to our active user community for finding some minor bugs in
our first implementation of UFO and for alerting us to actively used
extensions of the original UFO specification~\cite{Degrande:2011ua},
that were only later codified in UFO 2.0~\cite{Darme:2023jdn}.
Full detector simulation LCIO event files can now be fully
recast with {\sc Whizard}. Its scripting language, {\sc Sindarin},
now also allows for sum and product operator. There has also been a
validation of {\sc Whizard} with {\sc Pythia8}~\cite{Bierlich:2022pfr}
for parton showering and hadronization, where several issues had to be
fixed (cf. the talk by Z. Zhao at this conference). 
     

\section{Summary and Outlook}
\label{sec:outlook}

MC event generators are indispensable tools for High Energy
Physics~\cite{HEPSoftwareFoundation:2017ggl}. We report here on the
recent progress for the MC generator framework of {\sc Whizard}.
The two main general features published with the release series v.3.x
of {\sc Whizard} are the completion of the NLO automation for NLO SM
corrections at lepton and hadron colliders, as well as the completion
of the completely general UFO interface. Since the release of v3.0 in
April 2021, a few genuine bugs have been fixed in the UFO interface.
Support for important extensions that were unspecified
in~\cite{Degrande:2011ua} and codified in~\cite{Darme:2023jdn} has
been added. Ongoing work will enable the matrix-element generator of
{\sc Whizard}, {\sc O'Mega}, to provide matrix elements and their
squares exclusive in coupling orders, and will also support very
general color structures like vertices with epsilon tensors in color
space and color sextets or decuplets. An implementation of directly
providing displaced vertices with {\sc Whizard} is under way. Using
{\sc GoSam} as OLP, simple BSM extensions of the SM will be available
at NLO QCD, e.g. SMEFT. For precision simulations, there are several
ongoing projects, especially the validation and proper inclusion of
NLL electron QED PDFs for NLO EW calculations with massless leptons in
the inital state, the implementation of LL EW PDFs for high-energy
muon colliders, the implementation of YFS eikonal-based soft
resummation with exclusive photons and the implementation of a QED
parton shower for initial and final state. Regarding technical and
performance issues, work continues on the evaluation of matrix
elements on GPUs, an interface to HDF5 for grid files and event
formats work on the exclusive matched top threshold
implementation. Furthermore, several technical refactorings are in
progress.

\section*{Acknowledgments}

The work was partially supported by the National Science Centre
(Poland) under the OPUS research project no. 2021/43/B/ST2/01778.
KM and JRR acknowledge the support by the Deutsche
Forschungsgemeinschaft (DFG, German Research Foundation) under 
Germany's Excellence Strategy-EXC 2121 "Quantum Universe"-3908333 as
well as by the German Academic Exchange Service (DAAD) Cotutelle
Program -- 91837169. This research was supported by the Deutsche
Forschungsgemeinschaft (DFG, German Research Foundation) under grant
396021762 - TRR 257 as well as under grant 491245950.
TO is supported by the German Federal Ministry for Education and
Research (BMBF) under contract no.~05H21WWCAA.
We would also like to thank Christian Weiss for his continuous support,
and especially the GPU-based implementation for the matrix-element
evaluation and the profiling in {\sc Whizard}. We also like to thank
Maximilian Stahlhofen for his support on the vNRQCD top threshold
resummation. Our thanks also go to Marius H\"ofer for his support with
the {\sc GoSam} OLP, as well as to Jonas Lindert for the support with
{\sc OpenLoops}. For the NLO validation we also like to thank Stefan
Kallweit for many important cross checks of LHC processes at NLO. We
also like to thank Alan Price and Stefano Frixione for many
interesting discussions and explanations on YFS and collinear
factorization, respectively. For the validation of {\sc Whizard} in
the ILC setup, we gratefully acknowledge support from Mikael Berggren,
Jenny List, and Zhijie Zhao.


\baselineskip15pt
\bibliographystyle{utphys}
\bibliography{lcws2023_jrr}

\end{document}